# Structure of DNA-Functionalized Dendrimer Nanoparticles

Mattaparthi Venkata Satish Kumar and Prabal K Maiti[1]
Center for Condensed Matter Theory, Department of Physics, Indian Institute of Science, Bangalore 560012

## Abstract

Atomistic molecular dynamics simulations have been carried out to reveal the characteristic features of ethylenediamine (EDA) cored protonated (corresponding to neutral pH) poly amido amine (PAMAM) dendrimers of generation 3 (G3) and 4 (G4) that are functionalized with single stranded DNAs (ssDNAs). The four ssDNA strands that are attached via alkythiolate [-S $(CH_2)_6$-] linker molecule to the free amine groups on the surface of the PAMAM dendrimers observed to undergo a rapid conformational change during the 25 ns long simulation period. From the RMSD values of ssDNAs, we find relative stability in the case of purine rich (having more adenine and guanine) ssDNA strands than pyrimidine rich (thymine and cytosine) ssDNA strands. The degree of wrapping of ssDNA strands on the dendrimer molecule was found to be influenced by the charge ratio of DNA and the dendrimer. As G4 dendrimer contains relatively more positive charge than G3 dendrimer, we observe extensive wrapping of ssDNAs on the G4 dendrimer than G3 dendrimer. This might indicate that DNA functionalized G3 dendrimer is more suitable to construct higher order nanostructure. The linker molecule also found to undergo drastic conformational change during the simulation. During nanosecond long simulation some portion of the linker molecule was found to be lying nearly flat on the surface of the dendrimer molecule. The ssDNA strands along with the linkers are seen to penetrate the surface of the dendrimer molecule and approach closer to the center of the dendrimer indicating the soft sphere nature of the dendrimer molecule. The effective radius of DNA-functionalized dendrimer nanoparticle was found to be independent of base composition of ssDNAs and was observed to be around 19.5 Å and 22.4 Å when we used G3 and G4 PAMAM dendrimer as the core of the nanoparticle respectively. The observed effective radius of DNA-functionalized dendrimer molecule apparently indicates the significant shrinkage in the structure that has taken place in dendrimer, linker and DNA strands. As a whole our results describe the characteristic features of DNA-functionalized dendrimer nanoparticle and can be used as strong inputs to design effectively the DNA-dendrimer nanoparticle self-assembly for their active biological applications.

---

1    Email: maiti@physics.iisc.ernet.in

**Introduction**

In the emerging bio-nanotechnology field, the major focus is now on to induce biological functionality to organic or inorganic nanostructures. In the last 20 years, DNA has been considered to be a key component in bionanotechnology[1-5] and many novel materials with unique properties have been developed from DNA-linked nanoparticles,[6-9] polymers, or molecules. These DNA-functionalized nanoparticles have been demonstrated for their importance in wide range of applications, especially in molecular diagnostics. They can play important role in aggregation assays to detect single point mutations;[10] increase the sensitivity of micro-arrays,[11, 12] staining assays,[13] dipstick tests[14] and in biosensor applications.[15] In the recent past, the unique electrical, optical and catalytic properties of DNA-functionalized nanoparticles have been analysed for *in-vivo* applications such as drug delivery[16] and imaging.[17, 18] The recent discovery that crystalline rather than amorphous structures[19] can be produced when DNA links the particles together, leads to the development of new generation of DNA-linked materials chemistry. There exist also several other experimental[20-22] and theoretical studies[23] in the recent literature that have accounted for the DNA assisted crystallisation of colloidal nanoparticles. The properties of these crystalline materials are stated to depend on DNA length, number of DNA strands, rigidity of strand attachment, base pair sequence, shape and size of nanoparticle.[24-29] Several recent experimental and simulation studies have demonstrated the vast opportunities this field offers.[30]

In the recent years PAMAM dendrimers are considered to be one of the smartest building blocks to construct the nanostructure. Versatile applications of PAMAM dendrimers are made possible because of their exquisite control over size, shape, surface valency, and surface functionality in the nanoscale region.[31] Apart from these, PAMAM dendrimers were the first complete dendrimer family to be synthesized, characterized and commercialized[32, 33] and their derivatives were the most investigated experimentally.[34-42] In the class of PAMAM dendrimers, the different components are linked through covalent bonds by employing the charge interactions present in them. But this type of approach ends in difficulty because specific size ratios and covalent chemistry are necessary to coordinate the self-assembly of the core and shell dendrimers. Thus there is a need for the development of alternative method to assemble dendrimers into supramolecular clusters. In recent years DNA is found to be an ideal candidate for linking artificial constructs.[2, 43, 44] Base specificity and directional interactions make it possible to use it as a tool to self-assemble nanoscale components in precise structural arrangements.[45-47] So in this article we have attempted to study the DNA linked dendrimer nanoparticles which can be used to drive the self-assembly of the dendrimer nanoparticles.

The microscopic structure of DNA functionalized dendrimer nanoparticles (DNA-NPs) is not clear, and this is considered to be a hindrance in understanding how this system functions. Also to our knowledge the structure and properties of the DNA-functionalized and DNA-linked dendrimer nanoparticles have not been studied in details so far. Computer simulation has emerged as a powerful tool to study such problems. There are also simulation studies related to aggregation of PAMAM dendrimer,[48, 49] effect of explicit counter ions of the structure and dynamics of charged dendrimers,[50, 51] effects of PEGylation on the size and internal structure of PAMAM dendrimers,[52] as well as solution properties of amphiphilic dendrimers.[53] Sciortino et al.[54, 55] have used coarse grained model of DNA dendrimer and DNA functionalized nanoparticle to study their self-assembly. Recently, complexation of siRNA with dendrimer,[56] as well as the complexation between double-stranded DNA and various generations of PAMAM dendrimers[57] have been studied in great detail using all atom molecular dynamics simulations. It is believed that DNA functionalized dendrimer can be even better suited for the nucleic acid delivery materials as well as for driving the dendrimer self-assembly assisted by the DNA hybridization. This has motivated us to study the characteristic features of DNA-functionalized dendrimer nanoparticle using fully atomistic molecular dynamics simulations. Using classification scheme proposed by Tomalia[58-60] our DNA-NP system can be represented as [S-1 : ( S-6)4].

PAMAM dendrimers attached with ss-DNAs have not been synthesized yet. However, PAMAM dendrimers conjugated with fluorescein [FITC] and folic acid [FA] have been linked using complementary DNA strands for cancer therapy.[61] We hope that our study will inspire the synthesis of such system. Another important reason to study the structure of dendrimer nanoparticle linked to four ssDNA strands is the possibility for very unusual phase behaviour. Hsu and his co-workers[62] have explored the behaviour of an experimentally realized model for nanoparticles functionalized by four single strands of DNA and showed that this single-component model exhibits a rich phase diagram. Starr and Sciortino[63] have introduced and numerically studied a model designed to mimic the bulk behaviour of a system composed of single-stranded DNA dendrimers.

In this paper, using all atom molecular dynamics (MD) simulations we have studied eight different types of DNA-NPs at the atomistic level. Two different choices of dendrimers (G3 and G4) are considered as the core of the nanoparticle and four different choices of single stranded DNAs [ poly A strand $(-A-)_{10}$, poly C strand $(-C-)_{10}$, poly G strand $(-G-)_{10}$, and poly T strand $(-T-)_{10}$] are used to functionalize the nanoparticle. Below we give the details of the systems simulated and analyzed in this work:

1. Four poly(dA) strands attached to G3 dendrimer (system1)
2. Four poly(dC) strands attached to G3 dendrimer (system2)

3. Four poly(dG) strands attached to G3 dendrimer (system3)
4. Four poly(dT) strands attached to G3 dendrimer (system4)

Similarly four more following systems were simulated with G4 dendrimer as the core:
5. Four poly(dA) strands attached to G4 dendrimer (system5)
6. Four poly(dC) strands attached to G4 dendrimer (system6)
7. Four poly(dG) strands attached to G4 dendrimer (system7)
8. Four poly(dT) strands attached to G4 dendrimer (system8)

For the each of the above 8 types of DNA-linked dendrimer nanoparticle, we have performed 25 ns long MD simulations and studied the microscopic structure of the DNA-NPs. In these simulations, the ssDNA along with the linker is found to undergo fluctuation in the structure. The dendrimer (G3 and G4) also found to undergo structural changes during the course of simulation. Some of the ssDNAs along with linker were observed to penetrate the surface of the dendrimer and come close to the core of the dendrimer molecule. We also found some ssDNAs to wrap around the dendrimer and lose their helical structure. Earlier Lee et al.[64] studied the structural features of DNA-linked gold nanoparticles using atomistic simulation. There exist few theoretical models that are aimed at understanding the structural and thermal properties of materials in which DNA is used to link gold nanoparticles or polymers or organic molecules,[65] but this is the first simulation that has determined the characteristic structural features of DNA-NPs at the atomistic level.

Rest of the paper is organized as follows: In section 2 (a) we give the details of the building of the DNA-linked dendrimer systems, section 2 (b) gives the details of simulation. In section 3 we give details of the results from our atomistic simulation. Finally in section 4 we give a summary of the results and give an outlook for future possibilities.

## 2. Methods

**(a) Building of the system:**

In this study we have used protonated G3 and G4 PAMAM dendrimers. In the first case ($G_3 4A_{10}$) (as shown in **Figure 1A**), we have used protonated G3 dendrimer as the core of the nano particle and the four ssDNA strands, each composed of ten adenine bases, are linked to the four of the free amine groups on the dendrimer surface via a six-carbon alkylthiolate linker to functionalize the dendrimer. The initial structure of protonated G3 dendrimer was taken from our earlier studies.[36, 66-68] The four free amine groups on the dendrimer surface that are chosen to link the ssDNAs are symmetrically distributed around the periphery of the dendrimer particle. In the similar way the other seven systems $G_3C_{10}$, $G_3G_{10}$, $G_3T_{10}$, $G_4 4A_{10}$ (**as shown in Figure 1B**), $G_4C_{10}$,
4

$G_44G_{10}$ and $G_44T_{10}$ were built. The initial structure of ssDNAs containing ten adenine bases, ten cytosine bases, ten guanine bases and ten thymine bases are obtained using nucgen and xleap module of the Amber package.[69]

**(b) Simulation Methods:**

MD simulations are performed using PMEMD software package with the all-atom AMBER03 force field[70] for ssDNAs, GAFF for the linker alkylthiolate and Dreiding FF[71] for the G3 and G4 PAMAM dendrimer. The schematic representation of the six carbon alkylthiolate [-$S(CH_2)_6$-] linker molecule is shown in **Figure 1C**. The constrained RESP method was used to calculate the partial atomic charges of atoms in the linker molecule. The atom type, atom name and charge on each atom present in the linker molecule are shown in **Table 1**. The dendrimer structure of a given generation was connected to ssDNA strands via linker alkylthiolate by using the LEAP module in AMBER. The resulting structures were then solvated using TIP3P model of water[72] in the box of size 20 Å from the solute in the three directions. The negative charges on the ssDNAs and positive charges on the dendrimer of the system were neutralised by adding appropriate number of $Na^+$ and $Cl^-$ counter ions. Thus, we prepared the following systems: $G_34A_{10}$, $G_34C_{10}$, $G_34G_{10}$, $G_34T_{10}$, $G_44A_{10}$, $G_44C_{10}$, $G_44G_{10}$ and $G_44T_{10}$.

The resulting solvated structures were then subjected to 1000 steps of steepest descent minimization of the potential energy, followed by 2000 steps of conjugate gradient minimization. During the minimization, to remove the bad contacts between water molecules and the solute, the entire system (DNA-dendrimer) excluding waters were fixed in their starting conformations using harmonic constraints with a force constant of 500 kcal/mol/Å$^2$. The minimized structures were then subjected to 40 ps of MD using a 2 fs time step for integration. During the MD, the system was gradually heated from 0 to 300 K using 20 kcal/mol/Å$^2$ harmonic constraints on the solute to its starting structure. This allows for slow relaxation of the built DNA-dendrimer structure. In addition SHAKE constraints[73] using a geometrical tolerance of 5 x 10$^{-4}$ Å were imposed on all covalent bonds involving hydrogen atoms. Subsequently, MD was performed under constant pressure-constant temperature conditions (NPT), with temperature regulation achieved using the Berendsen weak coupling method[74] (0.5 ps time constant for heat bath coupling and 0.2 ps pressure relaxation time). This was followed by another 5000 steps of conjugate gradient minimization while decreasing the force constant of the harmonic restraints from 20 kcal/mol/Å$^2$ to zero in steps of 5 kcal/mol/Å$^2$. The above mentioned protocol produce very stable MD trajectory for such a complex system. Finally, for analysis of structures and properties, we carried out 25 ns of NVT MD using a heat bath coupling time constant of 1 ps. The analysis of structural parameters like RMSD, radial distribution function and radius of gyration are carried out using ptraj module of AMBER

package.[69]

## 3. Results and Discussion

**(a) RMSD Analysis:**

We have analysed the conformational changes in the ssDNAs during the time course of MD simulations. The initial built structure is used as a reference structure for the trajectory analysis. The fluctuation in the RMSD for the two ssDNAs of $G_34A_{10}$, $G_34C_{10}$, $G_44A_{10}$ and $G_44C_{10}$ systems as a function of time is shown in **Figure 2**. We observe similar trend in the RMSD fluctuation for the other two DNA strands of $G_34A_{10}$, $G_34C_{10}$, $G_44A_{10}$ and $G_44C_{10}$ systems (refer **Figure S1** in the supplementary section) and also for the four ssDNA strands of $G_34G_{10}$, $G_34T_{10}$, $G_44G_{10}$ and $G_44T_{10}$ systems (refer **Figure S2** in the supplementary section). In all the eight cases, the value of the RMSD of the ssDNA strands oscillates between 2 Å to 6 Å and in most of the cases the value settles down to a constant value after 6 ns. The average RMSD of each of four ssDNAs in all the eight systems after 15 ns simulation period is shown in **Table 2.** These average RMSD values are comparable with the values obtained for ssDNAs of similar length by Lee et al.[64] from their study on DNA functionalized gold nanoparticle.

From the average RMSD values and their standard deviation, it can be seen that the ssDNA strands in all the eight cases are found to undergo rapid conformational change during the course of simulation. This is mainly due to lengthening and shortening of ssDNA strands. We observe the RMSD values for some of the ssDNA strands to settle down to constant value in the initial period of the simulation and for some ssDNA strands the RMSD values are seen to fluctuate continuously during the entire time duration of simulation. These differences arise perhaps due to the position of the free amine group on the dendrimer surface to which the ssDNA strands are linked. Some of the free amine groups on the dendrimer surface (to which the ssDNA strands are linked) are flexible enough, thus those ssDNAs are observed to undergo rapid change in the conformation accordingly and thereby affects the structure of the dendrimer in turn. In this course of time the ssDNAs are found to lose their helical properties and wrap around the dendrimer surface (**as depicted in Figure 3 & 4**). And also the linker alkylthiolate group in between dendrimer and ssDNA strand was observed to move towards the dendrimer and lay almost flat (as shown in **Figure 3 & 4**) on the surface of the dendrimer. Thus there is very little contribution from the linkers to the radius of the DNA-dendrimers nanoparticle even though the length of the linker alkylthiolate is around 9 Å.
From the overall average RMSD values of ssDNA strands in $G_34A_{10}$, $G_34C_{10}$, $G_34G_{10}$, $G_34T_{10}$, $G_44A_{10}$, $G_44C_{10}$, $G_44G_{10}$ and $G_44T_{10,}$ we can infer the relatively better stability in the case of ssDNA strands that are composed of purine bases (adenine and guanine) than pyrimidine bases (thymine and cytosine). This may be due to better base-base stacking between purine bases than between pyrimidine bases. Mizutani et al.[75] also observed the presence of strong π-π stacking in solution and solid state of purine ring and not in pyrimidine ring.

**(b) End to end distance of the ssDNA strands**

The end to end distance of the four ssDNA strands during the time course of the simulation tell us about the degree of compactness and helical properties of the strands. In **Figure 5** we show the time evolution of the end to end distances for the two ssDNAs of $G_34A_{10}$, $G_34C_{10}$, $G_44A_{10}$ and $G_44C_{10}$ systems. We observe the similar end to end distance profile for the other two ssDNAs of $G_34A_{10}$, $G_34C_{10}$, $G_44A_{10}$, $G_44C_{10}$ systems (refer **Figure S3** in the supplementary section) as well as for the four ssDNAs of $G_34G_{10}$, $G_34T_{10}$, $G_44G_{10}$ and $G_44T_{10}$ systems (refer **Figure S4** in the supplementary section). The average length of each of the four ssDNAs in all the eight systems is shown in the **Table 3.** In all these cases, the length of the ssDNA strands are found to be oscillating rapidly over a wide range of values as the structure of ssDNA strand moves from collapsed conformation to the stretched conformation and vice versa. During this structural transition, some of the ssDNA strands are observed to lose their helical properties. The ssDNAs are found to wrap around the dendrimer surface due to electrostatic attraction between the positively charged amine groups and the negatively charged phosphate of the ssDNA. Similar wrapping pattern of ssDNA with G4 dendrimer was found in previous simulation studies.[76, 77] The length of ssDNAs is mainly affected due to its wrapping around the dendrimer surface. This wrapping is more when core of nano particle is G4 PAMAM dendrimer than G3 dendrimer. More positive charge on the G4 dendrimer surface compared to G3 dendrimer surface leads to extensive wrapping of ssDNAs on its surface. This can be seen from the instantaneous snapshots of the $G_34A_{10}$, $G_34C_{10}$, $G_34G_{10}$, $G_34T_{10}$ systems at the end of the 25 ns long MD run as shown in **Figure 6.** In **Figure 7** we show the snapshots for the $G_44A_{10}$, $G_44C_{10}$, $G_44G_{10}$ and $G_44T_{10}$ systems. Therefore to build the nanostructure with effective self-assembly of nanoparticles, it will be better to go for G3 dendrimer as the core of the nanoparticle than G4 dendrimer. As a whole the length of ssDNA strands reveal the extent of flexibility and structural changes in the strands. The tendency of ssDNA strands to wrap around G4 dendrimer can be minimized by using a double stranded segment of DNA near the surface of the core dendrimer or using a more rigid linker molecule.

**(c) Radial Distribution function**

To get the characteristic features of the DNA functionalized dendrimer structures, the radial distribution function of phosphate atom (of the first base of the ssDNA strand next to the alkylthiolate linker molecule) relative to the center of mass of the dendrimer molecule has been calculated at the beginning (0-1 ns) and at the end of the simulation period (24-25 ns). The results are depicted in **Figure 8** for $G_34A_{10}$, $G_34C_{10}$, $G_44A_{10}$ and $G_44C_{10}$ systems with separate curves for each of the two of the four DNA strands. We observe the radial distribution function profile for other four systems $G_34G_{10}$, $G_34T_{10}$, $G_44G_{10}$ and $G_44T_{10}$ to follow same trend (refer **Figure S5** in the supplementary section).

When we compare the radial distribution function of the phosphate atom relative to the center of mass of the dendrimer particle for all the eight cases at (0-1 ns) (shown in **Figure 8 and S5** as continuous lines) and at the end of the production period (i.e. at 24-25 ns) (shown in **Figure 8 and S5 as dashed lines**), we observe in most of the ssDNAs, a significant decrease in the distance between the center of mass of the dendrimer and the phosphate atom of the first base next to the linker alkylthiolate. During the simulation, the spatial orientation of linker molecule and ssDNAs changed to a great extent. Some part of the linker alkyl thiolate is found to lie almost parallel to the surface of the dendrimer. Lee et al.[64] in their study on DNA-gold structures, also observed the linker molecule to lie flat on the surface of the gold particle. And also it is seen that in some cases, the linker group along with the DNA strand has penetrated the surface of the dendrimer and moved closer to the center of mass of the dendrimer particle (**see Figure 6 & 7**). This in turn affected the shape and size of the dendrimer molecule in particular. The soft sphere nature of dendrimer can be inferred from this structural change. From these observations, we can appreciate the significant change in the structure of DNA-NP during the time course of simulation. All these microscopic pictures are very much useful to design and build nanostructure with effective self-assembly from the DNA-linked dendrimer molecule.

### (d) Radius of Gyration

In order to obtain the effective size of the DNA-NPs, the radius of gyration ($R_g$) for all the eight simulated systems were studied. The radius of gyration for the entire system and dendrimer alone excluding the linker, ssDNA strands are shown in **Figure 9A & B**. It is observed that the size of entire system and dendrimer alone in particular are not affected much with the different choices (compositions of bases) of ssDNAs. This is an important observation from our MD study.

Taking into consideration the fact that rise per base for double strand B-DNA is 3.4 Å and the length of the alkylthiolate linker group is about 9 Å. and the radius of gyration for G3 and G4 dendrimer to be around 17 Å and 22 Å respectively, the expected size of DNA-NP in the first four cases (where the G3 dendrimer is the core of the nanoparticle) and the last four cases (where the G4 dendrimer is the core of the nanoparticle) with a stretched conformation of DNA would be about 52.4 Å and 57.4 Å respectively. But from our simulation the value of $R_g$ (at 24-25ns) is found to be

19.95 ± 0.35 Å for $G_3 4A_{10}$, 19.32 ± 0.29 Å for $G_3 4C_{10}$, 19.46 ± 0.17 Å for $G_3 4G_{10}$, 19.83 ± 0.32 Å for $G_3 4T_{10}$, and 21.06 ± 0.16 Å for $G_4 4A_{10}$, 22.7 ± 0.27 Å for $G_4 4C_{10}$, 22.14 ± 0.3 Å for $G_4 4G_{10}$, 21.15 ± 0.23 Å for $G_4 4T_{10}$ .

There are many reasons that account for this difference. Part of the linker [-S (CH$_2$)$_6$-] group in most of the cases is found to lie flat on the surface of the dendrimer particle. As a result the contribution of linker group to the radius of the DNA-NP is greatly reduced. Lee et al.[64] in their study on DNA functionalized gold nanoparticle, observed no significant contribution from the linker -S(CH$_2$)$_6$- group to the radius of DNA-NP. In our case, the dendrimer particle acts as a soft sphere. Because of this, the linker along with DNA strands penetrate dendrimer surface and moved closer to the core of the dendrimer. It is also worth mentioning that in the all the eight cases, the free amine groups chosen on the surface of the dendrimer molecule where the ssDNA strands are connected via linker are not exactly at the equidistant position from the center of the dendrimer. So in some cases, some of the ssDNA strands were found to undergone rapid change in conformation and lost their helical structure, and finally end up in wrapping around the dendrimer particle (as shown in **Figure 3 & 4**). All these structural changes lead to shrinkage in size of the DNA-NPs during simulation (**see Figure 6 & 7**). Thus we observe lower $R_g$ values for our DNA-NPs. In the study of DNA-gold structures, Lee et al.[64] observed no ssDNA penetration into the gold nanoparticle and also the size of the gold nanoparticle was found to remain unchanged during the time course of the simulation period.

It is also seen that the size of DNA-NPs was found to be less significantly changed when we use G3 or G4 dendrimer as the core of the nanoparticle. The more positive charge on the G4 dendrimer surface resulted in the extensive wrapping of ssDNAs on its surface. So G3 dendrimer is preferred to G4 dendrimer as the core of the nanoparticle.

**(e) Density distribution**

The extent of penetration of DNA strands into the dendrimer surface and its resultant effect on the internal structure of the dendrimer in different complexes under study can be revealed by measuring and comparing the radial monomer density distribution ρ(r) of DNA, dendrimer and water. The radial monomer density profile for DNA, the dendrimer, and water for the $G_3 4A_{10}$, $G_3 4C_{10}$, $G_4 4A_{10}$ and $G_4 4C_{10}$ complexes is shown in **Figure 10**. We observe the similar drift in the density profile for the other four complex systems $G_3 4G_{10}$, $G_3 4T_{10}$, $G_4 4G_{10}$ and $G_4 4T_{10}$ (refer **Figure S6** in supplementary section). The radial monomer distribution has been calculated with respect to the center of mass of the dendrimer. From the density profiles, it is clearly seen that more and more DNA penetration has occurred in the case of G4 dendrimer compared to G3 dendrimer. Lee and Larson[52] in their study on the effect of PEGylation on the size and internal structure of dendrimers observed that longer chains and at higher concentrations of PEG self-penetrate inside dendrimer, occupying the dendrimer's vacant interior and result in a more dense shell dendrimer structure. We also see the swelling in dendrimers and a dip in dendrimer density in all the cases around 10 Å away from its core region. We observed more swelling in G4 dendrimers than G3 dendrimers. As a result penetration of large number of water molecules into the interior of G4 dendrimers is seen. This has actually lead to decrease in density and increase in size in the case of G4 dendrimer. So again we see advantage in G3 dendrimer than G4 dendrimer as the core of the nanoparticle.

4. **Conclusion**

In this work we have studied the characteristic features of DNA functionalized dendrimer nanoparticles at the atomistic level using MD simulations. The ssDNAs are found to lose their helical properties and wrap around the dendrimer surface. The extent of wrapping of ssDNAs on the dendrimer surface was observed to be dependent on the charge ratio of the dendrimer surface and ssDNA strand. Thus more wrapping was observed for G4 dendrimer surface as compared to the G3 dendrimer surface. This implies that G3 dendrimer may be more desirable building block than G4 dendrimer. We also observed ssDNAs containing purine bases to possess relatively more stability (in terms of RMSD) in the conformation than ssDNAs containing pyrimidine bases. Some portion of the linker molecule was found to orient nearly flat on the surface of the dendrimer molecule that has resulted in the lowering of the size of the DNA-functionalized dendrimer nanoparticle. In addition to this, radial distribution function of phosphate atom (of the first base next to the alkylthiolate linker molecule) relative to the center of mass of the dendrimer particle and also the density distribution function of DNA, dendrimer and water highlights the ssDNAs penetration at the surface of the dendrimer. We also observed the effective radius of DNA-NPs to be independent of composition of bases in the DNA strands. From this study, we have obtained strong inputs regarding the characteristic features of DNA functionalized dendrimer nanoparticle. We believe the microscopic picture of the DNA-dendrimer nanoparticle as revealed by our present simulation study will help designing the building blocks for various other DNA-linked nanoparticle systems and also to carry out the self-assembling process precisely. In future we plan to study the self-assembly of the such DNA functionalized dendrimers and see how various DNA sequence, length of DNA as well as the grafting density control the morphology of the resultant aggregates.

**Acknowledgements:**
We thank DST, India and MONAMI (Indo-EU project) for financial support.

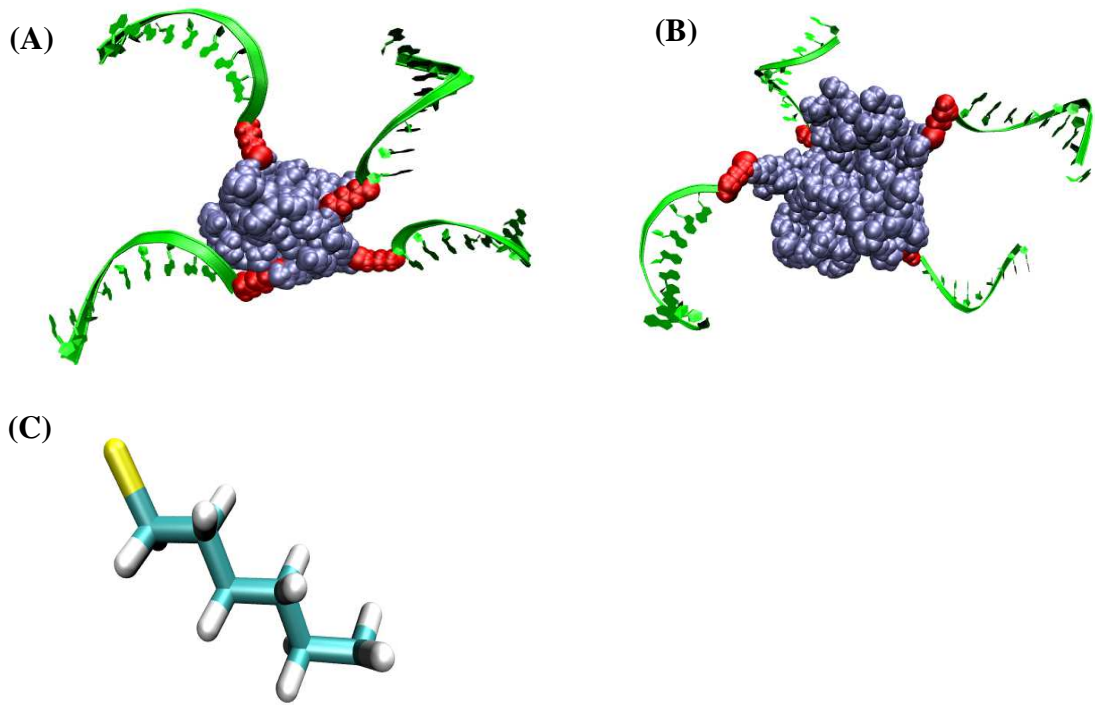

**Figure 1:** *(A) G3 dendrimer with four ssDNAs of 10 adenine bases ($G_3 4A_{10}$). (B) G4 dendrimer with four ssDNAs of 10 adenine bases ($G_4 4A_{10}$). Dendrimer and linkers are shown in the surface representation in iceblue and red color respectively, whereas ssDNAs are shown in the form of green ribbons. (C) Schematic structure of the six carbon alkylthiolate linker molecule [$-S(CH_2)_6-$]*

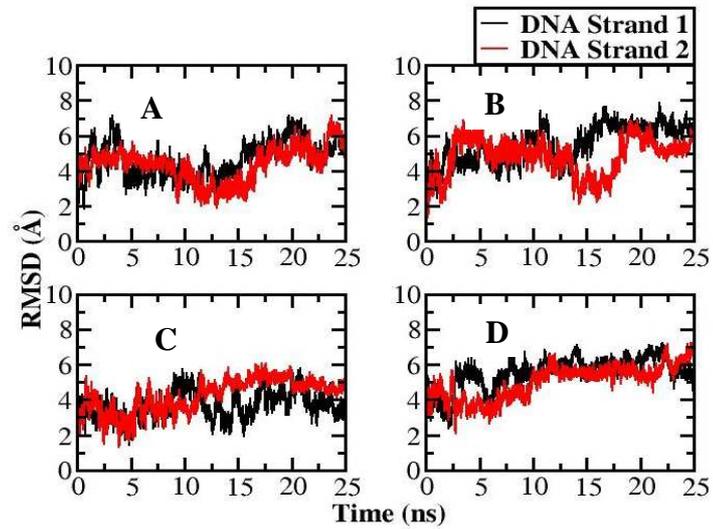

**Figure 2:** *Root mean square deviation (RMSD) of ssDNAs (strand 1 & 2) of (A) $G_3 4 A_{10}$ (B) $G_3 4 C_{10}$ (C) $G_4 4 A_{10}$ (D) $G_4 4 C_{10}$. In most of the cases the value of the RMSD becomes stable after 6 ns.*

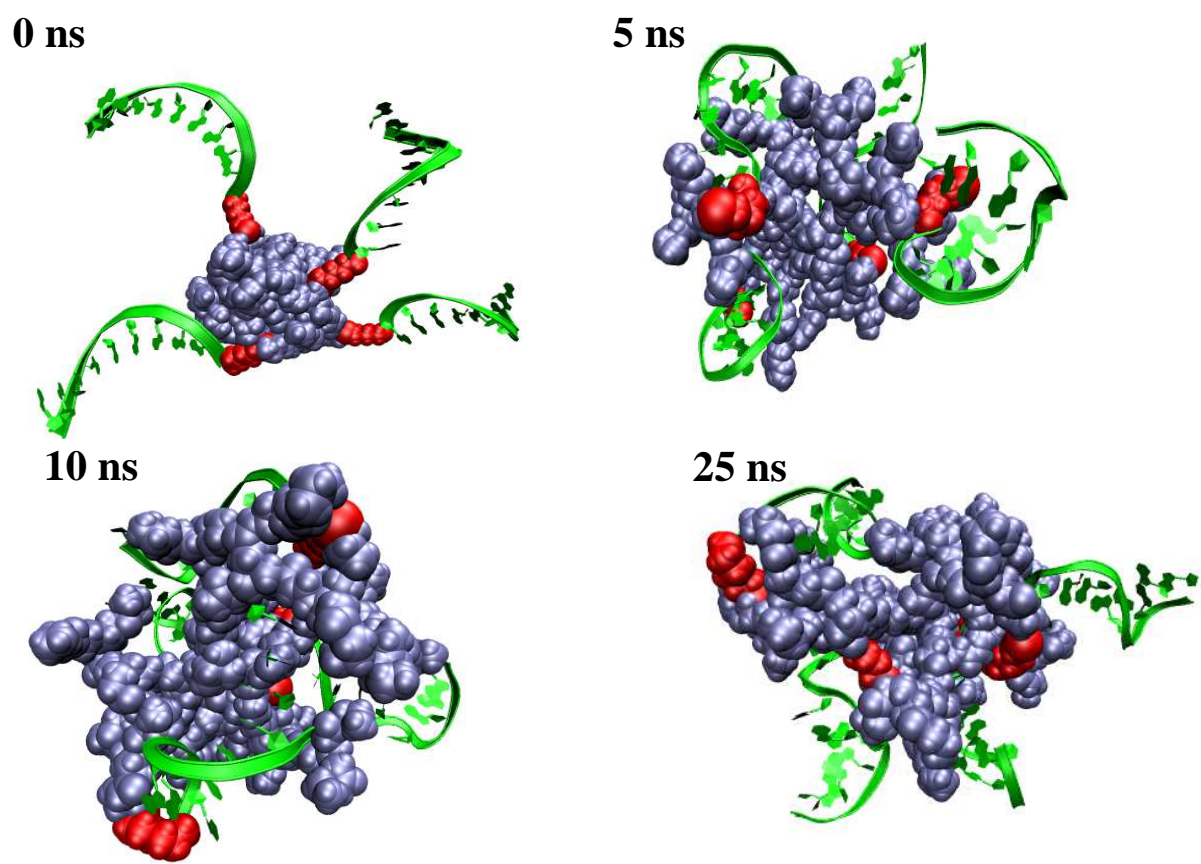

**Figure 3:** *Structure of $G_3A_{10}$ system during various stages of the wrapping process. Dendrimer and linkers are shown in the surface representation in iceblue and red respectively, whereas ssDNAs are shown in the form of green ribbons.*

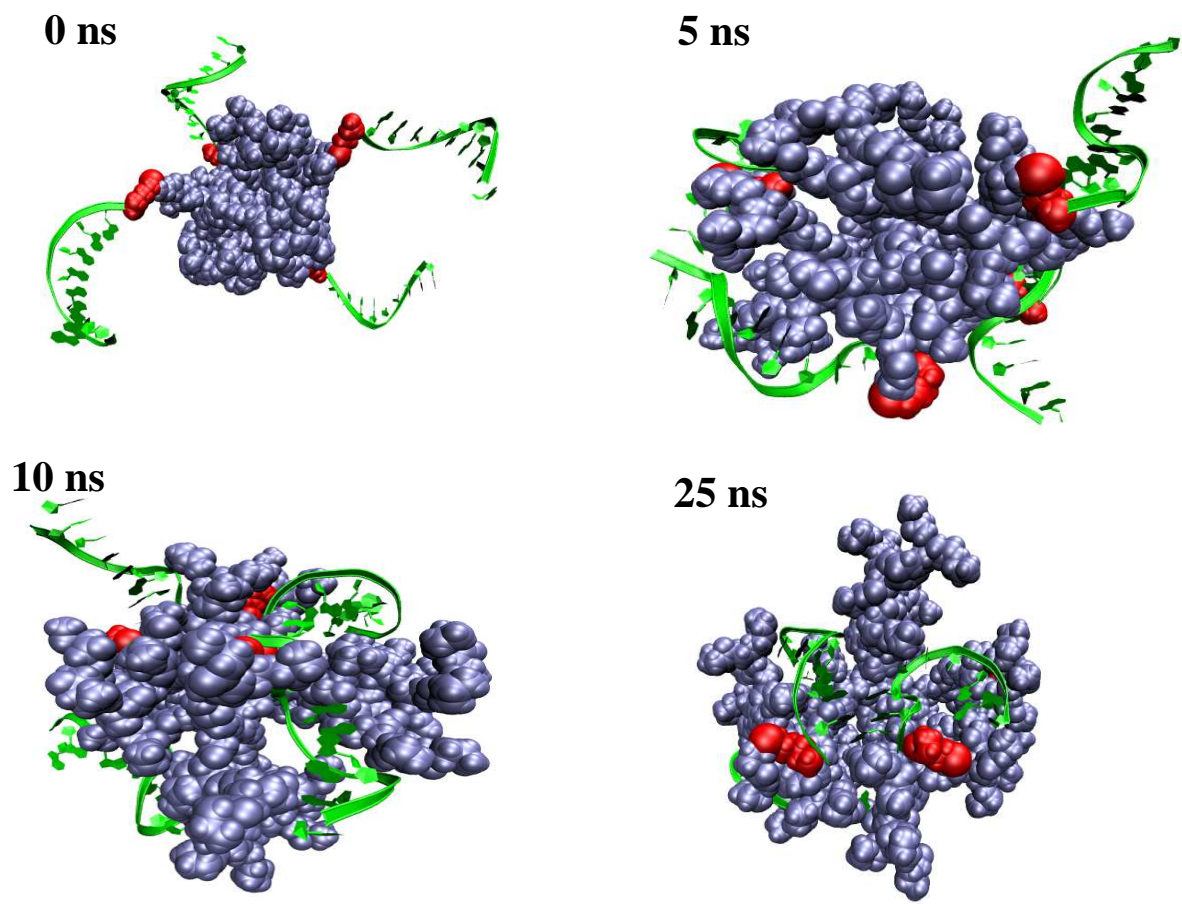

**Figure 4:** *Structure of $G_4A_{10}$ system during various stages of the wrapping process. Dendrimer and linkers are shown in the surface representation in ice blue and red respectively, whereas ssDNAs are shown in the form of green ribbons.*

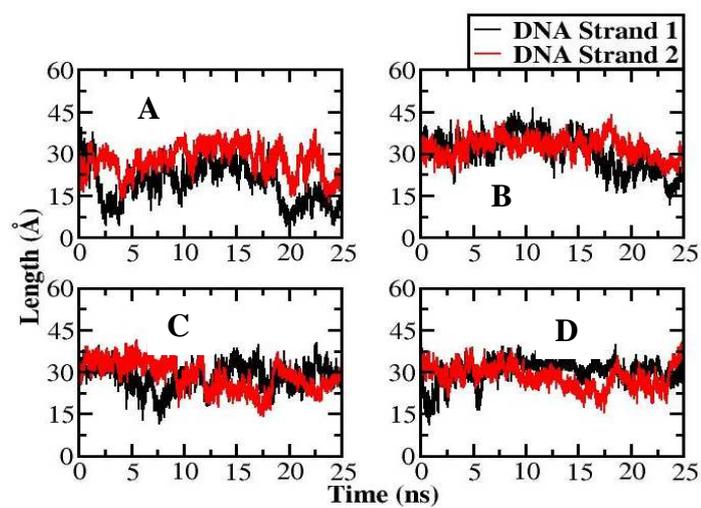

**Figure 5:** *End to end distance (Length) of ssDNAs (strand 1 & 2) of (A) $G_3 4A_{10}$ (B) $G_3 4C_{10}$ (C) $G_4 4A_{10}$ (D) $G_4 4C_{10}$.*

**(a) $G_34A_{10}$**

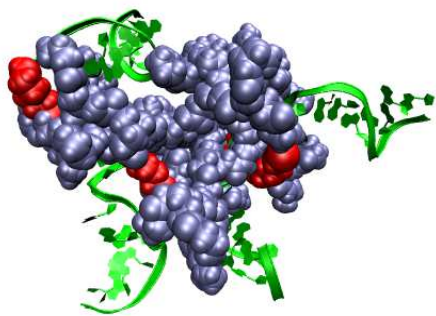

**(b) $G_34C_{10}$**

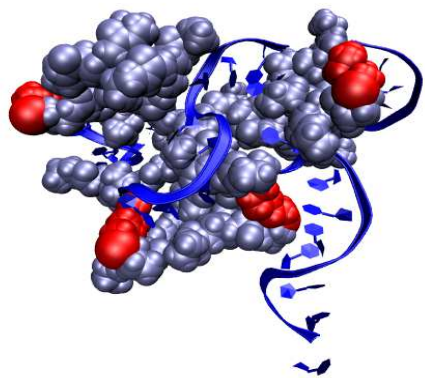

**(c) $G_34G_{10}$**

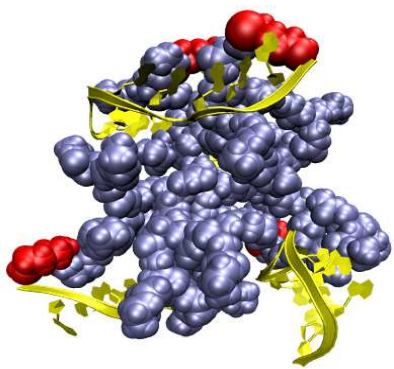

**(d) $G_34T_{10}$**

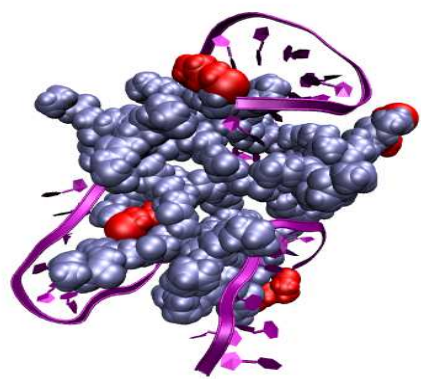

**Figure 6:** *25 ns equilibrated structure of (a) $G_34A_{10}$ (b) $G_34C_{10}$ (c) $G_34G_{10}$ (d) $G_34T_{10}$. Dendrimer and linkers are shown in the surface representation in iceblue and red respectively, whereas ssDNAs are shown in the form of green ribbons(for poly dA strands), blue ribbons(for poly dC strands), yellow ribbons(for poly dG strands) and purple ribbons(for poly dT strands).*

**(a) G$_4$4A$_{10}$**

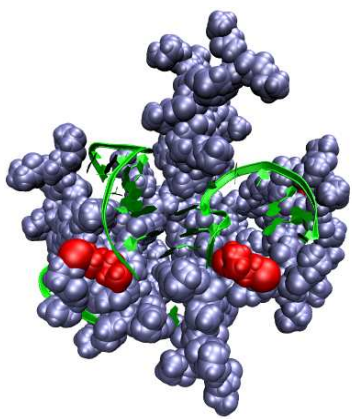

**(b) G$_4$4C$_{10}$**

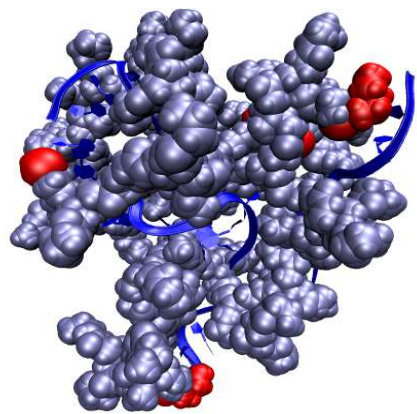

**(c) G$_4$4G$_{10}$**

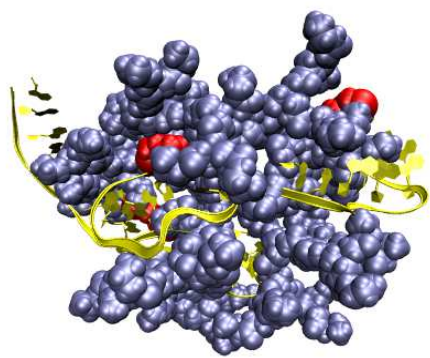

**(d) G$_4$4T$_{10}$**

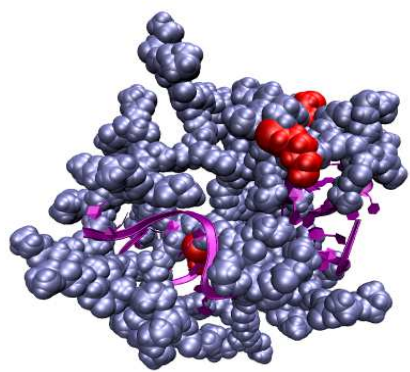

**Figure 7:** *25 ns equilibrated structure of (a) G$_4$4A$_{10}$ (b) G$_4$4C$_{10}$ (c) G$_4$4G$_{10}$ (d) G$_4$4T$_{10}$. Dendrimer and linkers are shown in the surface representation in iceblue and red respectively, whereas ssDNAs are shown in the form of green ribbons (for poly dA strands), blue ribbons (for poly dC strands), yellow ribbons (for poly dG strands) and purple ribbons (for poly dT strands).*

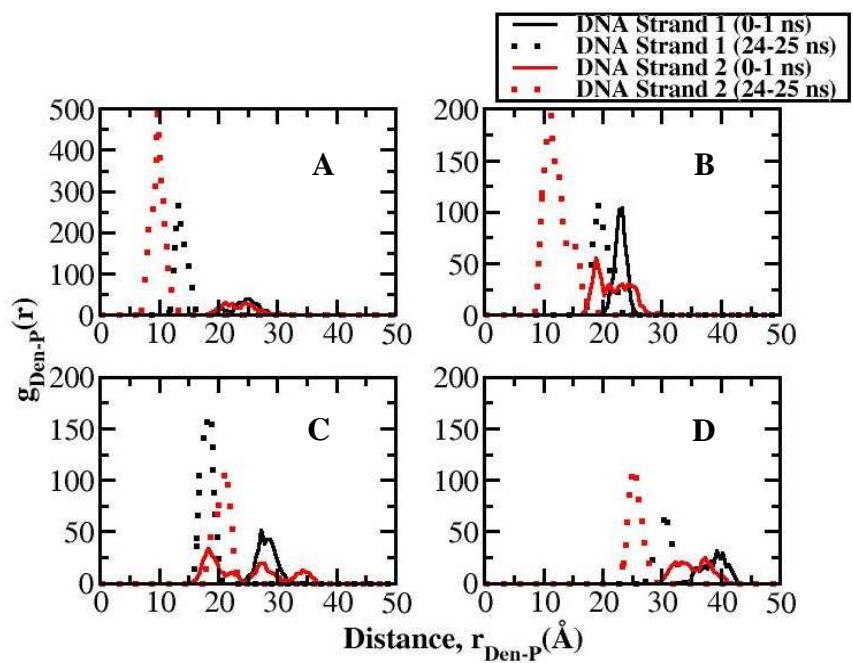

**Figure 8:** *Radial distribution function of the Phosphorous atom of the first base next to the linker molecule relative to the center of the dendrimer particle for (A) $G_3A_{10}$ (B) $G_3C_{10}$ (C) $G_4A_{10}$ (D) $G_4C_{10}$ at the (0-1) ns (represented using continuous lines) and at (24-25) ns (represented using dashed lines) of the simulation. Radial distribution function of each of ssDNA is shown separately in different colours.*

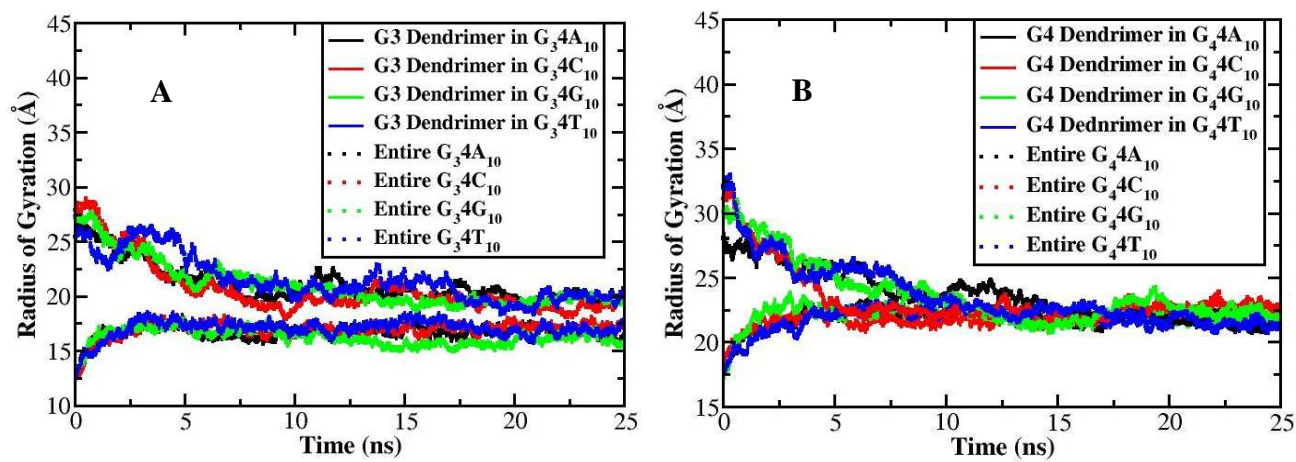

**Figure 9:** *Size of (A) G3 dendrimer and ssDNA linked G3 dendrimer (B) G4 dendrimer and ssDNA linked G4 dendrimer as a function of time.*

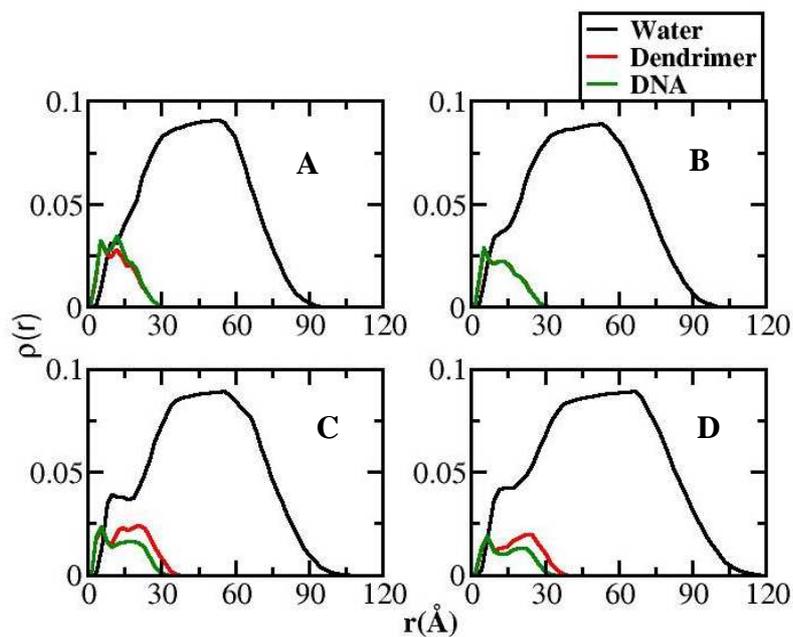

**Figure 10:** *Density distribution for DNA, the dendrimer, and water in (A) $G_3 4A_{10}$ (B) $G_3 4C_{10}$ (C) $G_4 4A_{10}$ (D) $G_4 4C_{10}$. The numbers shown were averaged from snapshots every 0.5 ps for the last 2 ns of the 25 ns long trajectory. The distribution has been calculated with respect to the center of mass of the dendrimer.*

**Table 1:** *Charges and atom types for the six carbon alkylthiolate linker molecule [-S (CH$_2$)$_6$-]*

| Atom name | Atom Type | Charge |
|---|---|---|
| C1 | c3 | 0.25326 |
| H1 | h1 | -0.02600 |
| H2 | h1 | -0.02600 |
| C2 | c3 | -0.05658 |
| H3 | hc | 0.00819 |
| H4 | hc | 0.00819 |
| C3 | c3 | 0.01926 |
| H5 | hc | -0.00414 |
| H6 | hc | -0.00414 |
| C4 | c3 | 0.82495 |
| H7 | hc | -0.18312 |
| H8 | hc | -0.18312 |
| C5 | c3 | -0.81641 |
| H9 | hc | 0.13847 |
| H10 | hc | 0.13847 |
| C6 | c3 | -0.54188 |
| H11 | h1 | 0.28904 |
| H12 | h1 | 0.28904 |
| S1 | ss | 0.51353 |

**Table 2:** *RMSD values of each of the four ssDNAs after 15 ns of simulation period in the different systems studied*

| System | DNA strand 1 | DNA strand 2 | DNA strand 3 | DNA strand 4 |
|---|---|---|---|---|
| **$G_34A_{10}$** | 5.45 ± 0.64 Å | 3.64 ± 0.63 Å | 3.89 ± 0.45 Å | 4.88 ± 0.93 Å |
| **$G_34C_{10}$** | 6.47 ± 0.40 Å | 4.8 ± 0.46 Å | 4.93 ± 1.06 Å | 4.16 ± 0.68 Å |
| **$G_34G_{10}$** | 3.97 ± 0.68 Å | 3.7 ± 0.38 Å | 5.6 ± 0.98 Å | 3.33 ± 0.61 Å |
| **$G_34T_{10}$** | 2.91 ± 0.50 Å | 4.21 ± 0.35 Å | 4.1 ± 0.51 Å | 5.37 ± 0.76 Å |
| **$G_44A_{10}$** | 5.97 ± 0.43 Å | 3.91 ± 0.67 Å | 4.99 ± 0.34 Å | 4.16 ± 0.35 Å |
| **$G_44C_{10}$** | 5.44 ± 0.89 Å | 6.03 ± 0.46 Å | 5.24 ± 0.59 Å | 5.73 ± 0.48 Å |
| **$G_44G_{10}$** | 4.11 ± 0.52 Å | 5.08 ± 1.34 Å | 6.12 ± 0.41 Å | 5.3 ± 0.27 Å |
| **$G_44T_{10}$** | 4.09 ± 0.49 Å | 4.58 ± 0.42 Å | 7.24 ± 0.33 Å | 6.13 ± 0.23 Å |

**Table 3:** Length *of each of the four ssDNAs in the different systems studied*

| System | DNA strand 1 | DNA strand 2 | DNA strand 3 | DNA strand 4 |
|---|---|---|---|---|
| $G_3 4A_{10}$ | 19.51 ± 7.2 Å | 32.56 ± 2.72 Å | 25.03 ± 6.89 Å | 27.62 ± 5.09 Å |
| $G_3 4C_{10}$ | 30.55 ± 6.54 Å | 32.51 ± 3.44 Å | 32.23 ± 3.57 Å | 27.38 ± 3.39 Å |
| $G_3 4G_{10}$ | 30.29 ± 6.9 Å | 27.24 ± 4.7 Å | 28.77 ± 5.24 Å | 30.2 ± 4.66 Å |
| $G_3 4T_{10}$ | 28.4 ± 5.58 Å | 27.45 ± 4.68 Å | 26.66 ± 5.76 Å | 34.11 ± 4.73 Å |
| $G_4 4A_{10}$ | 33.44 ± 3.91 Å | 28.08 ± 4.46 Å | 28.44 ± 5.14 Å | 23.62 ± 5.62 Å |
| $G_4 4C_{10}$ | 31.83 ± 5.52 Å | 29.98 ± 4.15 Å | 27.24 ± 4.45 Å | 28.23 ± 4.1 Å |
| $G_4 4G_{10}$ | 32.54 ± 2.68 Å | 32.37 ± 5.26 Å | 20.2 ± 8.49 Å | 28.27 ± 5.31 Å |
| $G_4 4T_{10}$ | 30.58 ± 3.09 Å | 22.92 ± 9.35 Å | 10.33 ± 9.07 Å | 17.61 ± 7.5 Å |